\pdfoutput=1
\documentclass[prl,showpacs,lengthcheck]{revtex4-1}
\usepackage{graphicx}
\usepackage{amsmath}
\usepackage{amssymb}

\newcommand{\ket}[1]{\ensuremath{|{#1}\rangle}}

\newcommand{\aat}{\ket{a}}
\newcommand{\bat}{\ket{b}}

\begin{document}

\title{Superfluidity of Interacting Bosonic Mixtures in Optical Lattices}

\author{Bryce Gadway}
\email{bgadway@ic.sunysb.edu}
\author{Daniel Pertot}
\author{Ren\'{e} Reimann}
\altaffiliation[Present address: ]{Institut f\"{u}r Angewandte Physik, Universit\"{a}t Bonn, $53115$ Bonn, Germany}
\author{Dominik Schneble}
\affiliation{Department of Physics and Astronomy, Stony Brook University, Stony Brook, NY 11794-3800, USA}
\date{\today}

\begin{abstract}
We report the observation of many-body interaction effects for a homonuclear bosonic mixture in a three-dimensional optical lattice with variable state dependence along one axis. Near the superfluid-to-Mott insulator transition for one component, we find that the presence of a second component can reduce the apparent superfluid coherence, most significantly when the second component either experiences a strongly localizing lattice potential or none at all. We examine this effect by varying the relative populations and lattice depths, and discuss the observed behavior in view of recent proposals for atomic-disorder and polaron-induced localization.
\end{abstract}

\pacs{03.75.-b, 67.85.Fg, 67.85.Hj, 63.20.K-}

\maketitle
Bosonic mixtures in optical lattices allow for the study of many interesting topics, such as the two-component Bose--Hubbard model~\cite{Kuklov-2003-SuperCounterFlowOL,*Altman-2003-TwoCompBHM,Isacsson-2005-TwoCompBHM,SoylerSansone-2009-SignAlter},
with its connection to quantum magnetism~\cite{Duan-2003-SpinSystemsOL,Kuklov-2003-SuperCounterFlowOL}, and models for decoherence mechanisms~\cite{OrthLeHur-2008-DissIsingSpinBoson}. In regard to condensed matter physics simulations, the introduction of a second component allows for the investigation of important phenomena such as polaron physics~\cite{Selftrapping1DSachaTimm-2006,Bruderer-2007-Polarons,Schirotzek-2009-FermiPolarons} and phonon-mediated long-range interactions~\cite{Bruderer-2007-Polarons,SoylerSansone-2009-SignAlter}, as well as effects of impurities and disorder~\cite{Gavish-2005-DisorderAtoms,RoscildeQuantumEmulsion-2007,Buonsante-2009-dBHMlocalimpurities}.

Recent experiments have addressed heteronuclear mixtures of atoms in optical lattices, both for the boson-boson~\cite{Catani-2008-KRbBoseBoseSFMI} and boson-fermion ~\cite{Gunter-2006-KRbBoseFermiSFMI,Ospelkaus-2006-KRbBoseFermiSFMI,BestWill-2009-KRbBoseFermiSFMI} cases. The superfluid coherence of the heavier bosonic component was universally found to decrease in the presence of the lighter second species, independent of the sign of the interaction, even for small overlap between the components~\cite{Catani-2008-KRbBoseBoseSFMI}. There exists a number of explanations for the observed behavior, ranging from localization due to impurities~\cite{Ospelkaus-2006-KRbBoseFermiSFMI},
self-trapping~\cite{BestWill-2009-KRbBoseFermiSFMI}, the formation of composite particles~\cite{Gunter-2006-KRbBoseFermiSFMI}, incoherent scattering of phonons~\cite{Catani-2008-KRbBoseBoseSFMI}, to thermalization effects~\cite{Gunter-2006-KRbBoseFermiSFMI} and changes in the chemical potential~\cite{Buonsante-2008-BoseBoseMix}. Compared to the single-component case, the description of mixtures presents richer physics, but also depends on additional parameters, such as the ratio of the tunneling rates, the interspecies interaction, and the relative atom numbers.

In this paper we study how the superfluid coherence of bosons in a lattice is affected by a variable bosonic ``background'' medium. Through a state-dependent optical lattice we can vary the localization of the medium's constituent atoms, and we find a non-monotonic dependence of the coherence properties of the primary species (``foreground atoms'') on this localization. For a binary mixture of $^{87}$Rb hyperfine states, we demonstrate a reversible state-dependent transition from the superfluid into the Mott regime, and we systematically examine interaction effects by varying both the relative populations and the respective lattice depths. Remarkably, the superfluid coherence decreases not only when the bosons of the medium are strongly localized, but also when they are very delocalized. We present indirect evidence that this may be due to polaron-type dynamics in the optical lattice.

Our experimental setup has been described in detail in
Ref.~\cite{Pertot-09-JPB}. In brief, we produce a $^{87}$Rb~Bose--Einstein
condensate in the $\ket{F,m_F} = \ket{1,-1}$ hyperfine ground state
containing $3-5\times10^5$ atoms in a crossed-beam optical dipole trap (ODT)
at $\lambda_\perp = 1064$~nm, with a mean trap frequency of about 50~Hz. We prepare mixtures of the $\ket{1,-1}\equiv\aat$
and $\ket{2,-2}\equiv\bat$ hyperfine ground states, with variable fractional
populations $f_a$ and $f_b=1-f_a$, via microwave Landau--Zener
sweeps~\cite{Mewes-1997-LandauZener}. These mixtures are subsequently loaded into an
adiabatically ramped up 3D optical lattice. The state-independent transverse lattice potentials along $x$ and $y$ are generated by partial retroreflection of the ODT beams~\cite{Pertot-09-JPB}, maintaining constant
gravitational sag through coordination of the forward beam intensities with
the amount of retroreflection. The state-dependent lattice along $z$ is generated by full
retroreflection of a beam at $\lambda_z=785.1$~nm ($1/e^2$ radius:
$230~\mu$m). At this wavelength (between the D$_1$ and D$_2$ lines of $^{87}$Rb),
the individual depths $V_a$ and $V_b$ of the $z$-lattice, for $\aat$ and
$\bat$, are strongly polarization
dependent~\cite{DeutschJessen-1998-QuantStateControlOptLat}.
A $0.4$~G magnetic bias field along $z$ defines the quantization axis. By
varying the polarization of the beam from $\sigma^+$ to $\sigma^-$, we can adjust the ratio $V_b/V_a$ from
approximately 0 to 3.5. We characterize the many-body state in the lattice by releasing the atoms and then absorptively imaging both components on the $F=2\to F'=3$ cycling transition after 18~ms time-of-flight (TOF), concurrent with optical pumping from $F=1\to F'=2$. Additionally, we use a magnetic Stern--Gerlach pulse to separate the states in TOF.

The use of a homonuclear mixture avoids differential gravitational sag~\cite{Catani-2008-KRbBoseBoseSFMI} in our far-detuned ODT, and the similarity of all relevant scattering lengths (difference $<2$\% \cite{Kokkelmans-PC10}) precludes macroscopic phase separation. We characterize component overlap in the ODT using collinear two-component four-wave mixing as a sensitive probe~\cite{Pertot-09-FWM}. Residual magnetic field gradients were carefully canceled by maximizing component overlap and inferred to be less than $40~\mu\mbox{G}/\mbox{cm}$ from hyperfine Ramsey measurements.

\begin{figure}[t!]
\centering
   \includegraphics[width=0.9\columnwidth]{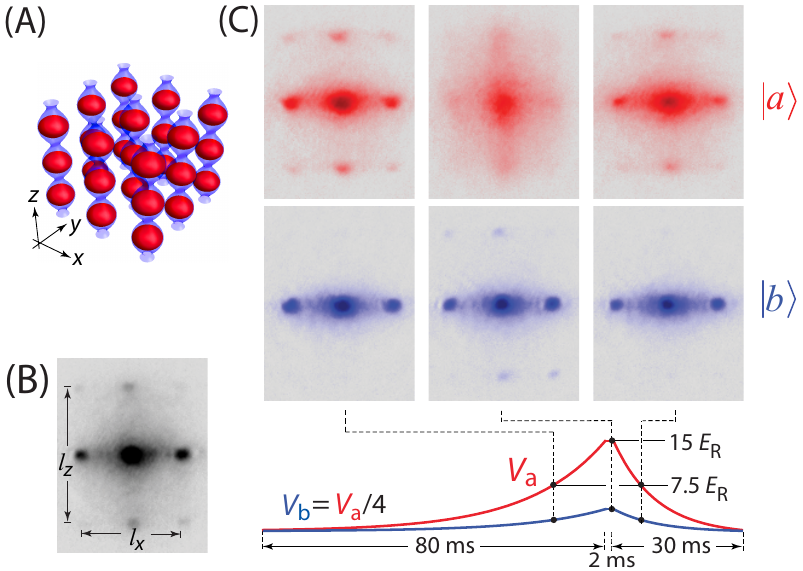}
    \caption{(Color online) State-dependent transition from the superfluid to the Mott regime. (A) Surfaces of equal probability density ($1/e^2$) for (non-interacting) atoms of type $\aat$ (red) and $\bat$ (blue) in a three-dimensional lattice with depths of 12~$E_R^\perp$ in the transverse directions ($x$, $y$), and 15~(3.8)~$E_R$ along $z$ for $\aat$ ($\bat$) atoms. (B) Time-of-flight (TOF) absorption image after release of a balanced mixture ($f_a \approx f_b$). The aspect ratio $l_z/l_x$ is given by $k_z/k_\perp$. (C) TOF images after Stern--Gerlach separation. The $\aat$ component (red) enters the Mott regime (with $t_{a}/U_{aa}\approx1/39$, and $t_\perp/U_{aa}\approx1/38$), whereas the $\bat$ component (blue) remains superfluid. The bottom graph illustrates the state-dependent $z$-lattice ramp. The transverse lattice ramp (not shown) follows a sigmoid curve of 115~ms duration and reaches its full depth of $12 \ E_R^\perp$ as the $z$-lattice goes through half of its maximum depth.}
    \label{FIG:TransImg}
\end{figure}

We first demonstrate a state-dependent transition from the superfluid to the Mott regime, keeping the final transverse lattice depth constant at 12~$E_R^\perp$ ($E_R^\perp = \hbar^2 k_\perp^2/2m$, $k_\perp = 2\pi/\lambda_\perp$) and ramping up and down the $z$-lattice depth. The transverse lattice depth is slightly below the superfluid-to-Mott insulator transition in an isotropic 3D cubic lattice (at a tunneling-to-interaction ratio $t/U \approx 1/36$~\cite{Greiner-SF-MI-2002}), and sufficiently large for unwanted four-wave mixing effects in TOF~\cite{Pertot-09-FWM} to be negligible. Fig.~\ref{FIG:TransImg} shows TOF images for the case $V_a=4 V_b$ and a ramp of $V_a$ up to 15~$E_R$ ($E_R = \hbar^2 k_z^2/2m$, $k_z = 2\pi/\lambda_z$). The $\aat$ atoms undergo a reversible transition into the Mott regime, while the $\bat$ component remains superfluid.

\begin{figure}[t!]
\centering
    \includegraphics[width=\columnwidth,clip=true]{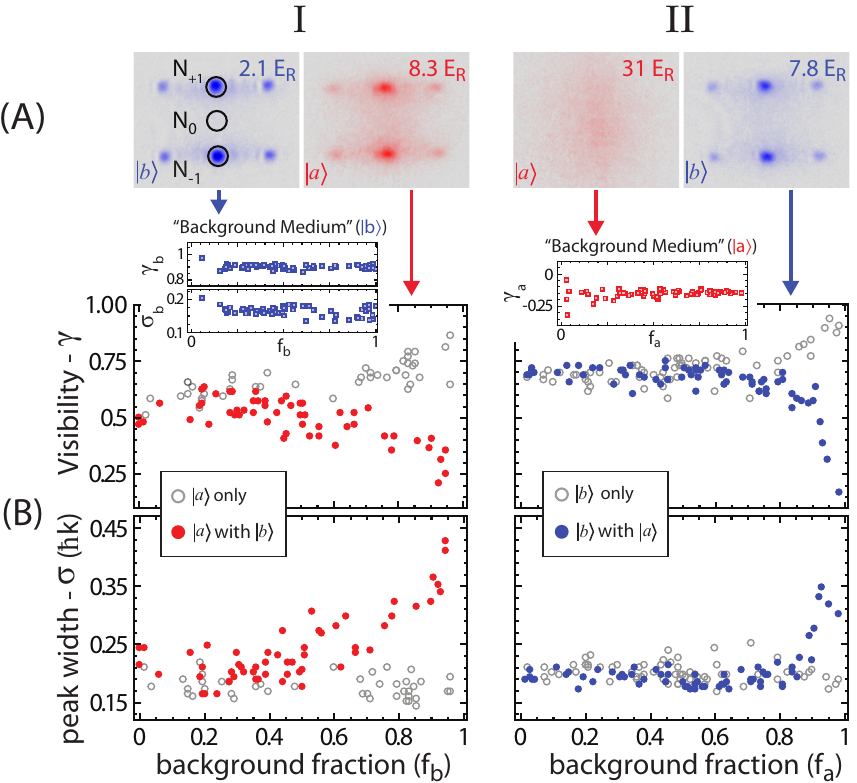}
    \caption{(Color online) Effects of a background medium on superfluid coherence.  (I;left column) $\aat$ atoms ($V_a = 8.3~E_R$) in contact with superfluid  $\bat$ atoms ($V_{b}=2.1~E_R$) and (II;right column) $\bat$ atoms ($V_b = 7.8~E_R$) in contact with localized $\aat$ atoms ($V_{b}=31~E_R$). (A) Stern--Gerlach separated TOF images (for $f_a \approx f_b$) after a gravitational phase shift. The circles in (I) denote apertures used to determine visibility. (B) Dependence of the visibility and peak width on the relative population of the background medium, with constant total atom number $1.7(1) \times 10^{5}$; the insets show corresponding data for the background medium. The open circles denote reference measurements without background medium, in which the atom numbers are varied correspondingly.  All ramp shapes are as in Fig.~\ref{FIG:TransImg}, with $V_{\perp}=12~E_R^\perp$.}
    \label{FIG:VisPlot}
\end{figure}

To characterize the many-body states of the two components, we analyze the
visibility~\cite{Gerbier-2005-MottInsVisibilityPRL} and peak width~\cite{Kollath-2004-PeakWidth} of their TOF diffraction patterns. As our optical lattice is anisotropic (transverse and vertical axes) in both
lattice constant and depth, we employ a single-axis visibility measure along the state-dependent vertical axis. A
600$~\mu$s gravitational $\pi$-phase shift~\cite{Orzel-2001Squeezed,Greiner-SF-MI-2002} between adjacent vertical
lattice sites produces a symmetric diffraction
pattern with two vertical peaks separated by $2\hbar k_z$ (in contrast to the patterns of Fig.~\ref{FIG:TransImg} without a shift). This is accomplished by turning off the ODT while keeping on the $z$-lattice. The visibility is then defined as $\gamma=N_-/N_+$, where $N_{\pm} = N_{+1} + N_{-1} \pm 2N_0$ are the sum and difference of the atom numbers in the diffraction
peaks ($N_{+1},N_{-1}$) and intermediate region ($N_0$), as shown in
Fig.~\ref{FIG:VisPlot}~(I-A). To determine the peak width $\sigma$ ($1/e$ half-width), we sum over a vertical strip and fit the projection with two Gaussian peaks on top of a broad Gaussian background.

In the following, we examine how the coherence properties of a given superfluid component depend on the presence of a background of either a lighter (in terms of the band mass $\textit{m}^*_z$) superfluid, or heavier ``impurity'' atoms. This is implemented by a halving (I) and doubling (II) of the maximum $z$-lattice depths as compared to Fig.~\ref{FIG:TransImg}. In the first case (I) the background medium is formed by $\bat$ atoms in a $2 \ E_R$ deep potential, and in the second case (II) by $\aat$ atoms in a potential with $31 \ E_R$ depth. In both cases, the foreground component ($\aat$ atoms in (I), $\bat$ atoms in (II)) experiences an $8 \ E_R$ deep lattice, for which the visibility displays a strong differential dependence on lattice depth and tunneling (at the chosen transverse lattice depth of 12~$E_R^\perp$). In both cases, we vary the relative populations $f_a$ and $f_b = 1-f_a$ of the two components while keeping the total atom number constant, which allows for separating out effects of interspecies coupling from simple overall density effects.

In case (I) the background density is only weakly modulated by the $2 \ E_r$ deep $z$-lattice, while the foreground atoms are concentrated on lattice sites to a good approximation, but can still tunnel appreciably. The collisional interaction will thus tend to repel background atoms away from sites populated by the foreground atoms. The resulting dips in the background density (within a characteristic range given by the mean-field healing length $\xi$) in turn increase the foreground atoms' localization. This mechanism, similar to self-trapping of impurities in a condensate~\cite{Selftrapping1DSachaTimm-2006}, corresponds to the formation of polarons~\cite{Bruderer-2007-Polarons} - composite quasi-particles consisting of individual foreground atoms surrounded by a coherent phonon cloud comprising the background's local mean-field depression. The immersed foreground atoms should thus have an increased effective mass and a lower tunneling than bare foreground atoms, leading to a degradation of the foreground's observed coherence.

Data for case (I) are shown in Fig.~\ref{FIG:VisPlot}~(I). As the superfluid background of $\bat$ atoms gets more and more populated at the expense of the foreground $\aat$ component, the $\aat$ visibility drops continuously and the diffraction peaks get broader. A reference without background, in which the $\aat$ atom number is correspondingly varied, in contrast reveals a slight increase in visibility with a reduction in $\aat$ atom number, consistent with the naive expectation for a decrease in on-site density. We note that atom transfer from the $\aat$ component to the less localized $\bat$ component lowers the on-site density as well, which naively should also lead to an increase in visibility. The observations thus are not consistent with a simple overall density effect. Additionally, the visibility could in principle be affected by incoherent intra- and interspecies collisions after release from the trap. However, both would similarly degrade the $\bat$ visibility which, as seen in the inset of Fig.~\ref{FIG:VisPlot}~(I-B), remains at a consistently high value throughout.

The observed reduction of the $\aat$ visibility from $\gamma_a \approx 0.5$ (no atoms in background) to $\gamma_a \approx 0.3$ (95\% in background), and the associated increase in peak width, would correspond in the single-component case to a reduction of tunneling by about 40\%, as estimated from an independent single-component reference measurement in which the lattice depth was varied. For a single immersed $\aat$ atom, the surrounding dip of the $\bat$ component density leads to an on-site polaronic energy shift $V_p = g_{ab,1\mathrm{D}} / \xi\sim 0.2~E_R$~\cite{Bruderer-2007-Polarons} (where $\xi\sim 200$~nm, $g_{ab,1\mathrm{D}} = 2 a_{ab} \hbar \omega_{\perp}$ and $a_{ab}\approx a_{bb}\approx 100a_0$ is the interspecies scattering length), and one can roughly estimate that for the modified on-site potential depth $\sim V_a + V_p$ there is an approximately 5\% reduction in tunneling. Therefore, this single-polaron effect alone seems insufficient to explain the observed change in visibility. However, given the large number of $\aat$ atoms even at $f_b = 0.95$, additional localization can be expected due to polaron clustering~\cite{Bruderer-2007-Polarons}, which results from attractive off-site interactions mediated by the superfluid background~\cite{SoylerSansone-2009-SignAlter}. The mutual exponential localization of polarons in clusters has been predicted to lead to a broadening of the momentum distribution~\cite{Bruderer-2007-Polarons}, similar to our observations.

With regard to possible temperature effects, adiabatic loading of bosons into the lowest band of a 3D optical lattice should lead to lower temperatures due to the reduction in band width~\cite{Blakie-2004-LatticeCooling}. Given the unequal $z$-lattice depths, one thus would expect unequal final temperatures for the two components if loaded separately, leading to thermalization in a mixture~\cite{McKay-2009}. However, thermalizing elastic collisions should be largely suppressed due to the mismatch in band structure~\cite{Pertot-09-FWM,McKay-2009}.
Moreover, the expected temperatures in the lattice are incompatible with the magnitude of the observed effects, even assuming thermalization. Reference measurements with only $\aat$ atoms showed that visibilities below $0.3$ require initial loading temperatures exceeding 75~nK, which can be estimated \footnote{using $T_f / T_i \sim \textit{m}/\textit{m}^*$~\cite{Blakie-2004-LatticeCooling}, with $\textit{m}^*=\sqrt[3]{\textit{m}^*_x \textit{m}^*_y \textit{m}^*_z}$ for our anisotropic lattice~\cite{BlakieWang-BECopticallattice-PRA2007}} to yield final temperatures above $14$~nK. However, for the mixture we start with much lower initial loading temperatures $\sim$15~nK that should accordingly be reduced to below 4~nK even in the weaker $\bat$ lattice (also, these temperatures $\ll V_p/k_B$ are sufficiently low to suppress effects of thermal phonons on polaron tunneling~\cite{Bruderer-2007-Polarons}).
We can also exclude trivial heating effects due to spontaneous photon scattering from the state-dependent lattice beam, since the $\bat$ atoms are not only less confined to intensity maxima, but also have a slightly lower scattering rate, so that such heating should be lower for higher values of $f_b$.

We now turn to the case (II) of a superfluid coupled to a background medium that is pinned to the optical lattice. We ramp the $z$-lattice to $V_{b (a)} \approx 8 (31) \ E_R$, for which the $\aat$ component, now playing the role of the background, is deep in the Mott regime. With an increasing background fraction, we again observe a drop in visibility and an increase in peak width of the foreground $\bat$ atoms, as is shown in Fig.~\ref{FIG:VisPlot}~(II).

The observed changes for the $\bat$ component could be caused by a simple increase of on-site interactions due to localized $\aat$ atoms. However, the changes to both visibility and peak width are quite abrupt, which may be indicative of a disorder effect. The simultaneous loading of multi-component gases with differing lattice parameters is predicted to result in a ``quantum emulsion state''~\cite{RoscildeQuantumEmulsion-2007,Buonsante-2008-BoseBoseMix} similar to a Bose-glass, with the less mobile species acting as quasi-static impurities. The observed sudden drop could thus be caused by the concentration of localized $\aat$ atoms increasing beyond a percolation threshold, leading to localization of the $\bat$ atoms~\cite{Buonsante-2009-dBHMlocalimpurities}. Another possible contribution, heating due to spontaneous photon scattering by $\aat$($\bat$) atoms, is measured to amount to less than $14(8)$~nK for symmetric ramps to $31(8)~E_R$ and back down. However, thermalization (which by the earlier argument should be largely suppressed due to disparate band structures) would not appear to be consistent with the suddenness of the change in $\bat$ visibility and peak width at $f_a \sim 0.8$ for a continuous increase of background $\aat$ population.

\begin{figure}[t!]
\centering
   \includegraphics[width=0.7\columnwidth,clip=true]{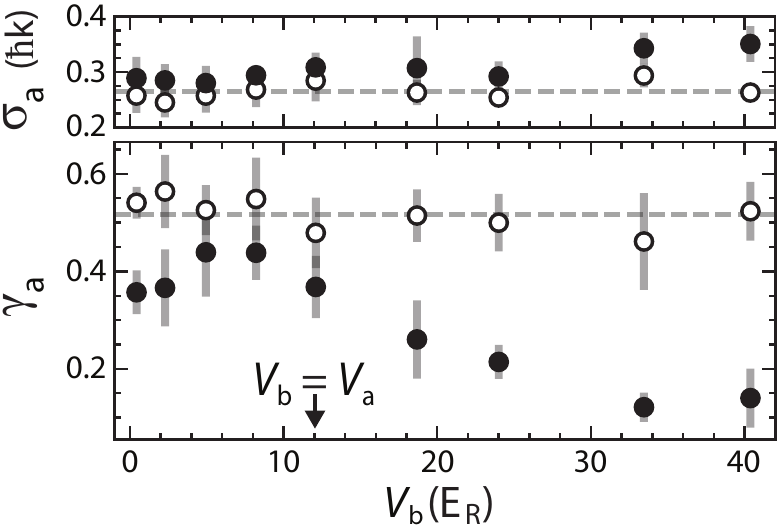}
    \caption{Dependence of superfluid coherence on the localization of the second component. Visibility ($\gamma_a$) and peak width ($\sigma_a$) (filled circles) of the $|a\rangle$ component, with fixed lattice parameters of $V_a = 12 \ E_R$ and $V_{\perp}=12 \ E_R^\perp$, in the presence of $\bat$ atoms ($f_b \approx 3/4$,\ $N_a + N_b = 3.7(2) \times 10^{5}$) as $V_b$ is increased. In references (open circles) taken without $\bat$ atoms and with $N_a = 1.0(1) \times 10^{5}$, the visibility and peak width are roughly constant at $0.52(3)$ and $0.26(2)$ (dashed lines). Data points are averaged over 3-5 runs, with statistical error bars shown. Uncertainties in lattice depth are $\sim$5\% (not shown).}
    \label{FIG:VaryingPolarization}
\end{figure}

Finally, having studied these two extreme cases, we examine more closely how the interspecies effects depend on the background medium's degree of localization at a fixed background fraction. Using the polarization dependence of our lattice, we keep $V_a$ fixed at $12~E_R$ while tuning $V_b$ over a wide range, changing the character of the $\bat$ background component from superfluid to localized. Shown in Fig.~\ref{FIG:VaryingPolarization} are results for a mixture with $f_b \approx 3/4$. The addition of $\bat$ atoms leads to a reduction of the $\aat$ visibility, most prominently when the $\bat$ component is either much lighter or much heavier, in terms of band mass. For $V_b = V_a$, the visibility is lower than in a reference measurement in which only $\aat$ atoms were present, which is most likely caused by the near doubling of the chemical potential due to the increase in overall atom number.
We note that the strongest interspecies effects are observed when the background $\bat$ atoms are highly localized, while no effect is observed for a similar background fraction of localized atoms in Fig.~\ref{FIG:VisPlot}~(II). However, here we have a much greater total population (more than twice that of Fig.~\ref{FIG:VisPlot}~(II)), so that the concentration of sites occupied by localized background atoms will be greater for a given background fraction, which should enhance the disorder-induced localization of the foreground component as discussed above.
In general, the foreground visibility at a given lattice depth should depend on the modified tunneling and on-site interactions due to background atoms, as well as non-trivial effects of clustering and disorder as discussed above. Thus, a determination of where the visibility maximum is located would require a quantitative model for the detailed interplay between these effects.

In summary, we have studied interspecies effects for a binary bosonic mixture, in a 3D optical lattice with tunable state dependence along one axis. We observe a reduction of apparent superfluid coherence, most strongly for large population imbalances and tunneling rate asymmetries of the two components. The observed reduction in coherence for the addition of both a delocalized and also a localized second species suggest polaron-related effects and atomic-disorder, respectively. Our system should be of interest for future investigations of phonon-mediated interactions in polaron systems and in the spin-boson model.

\begin{acknowledgments}
We thank B. Capogrosso-Sansone and W. Ketterle for discussions and valuable comments, and B. Bogucki for important contributions to the apparatus. This work was supported by NSF (PHY-0855643), ONR (DURIP), and the Research Foundation of SUNY. B.G.\ acknowledges support from the GAANN program of the US DoEd.
\end{acknowledgments}

\end{document}